\documentclass[aps,pra,preprint,groupedaddress,showpacs]{revtex4-1}
\usepackage{graphicx}
\usepackage{dcolumn}
\usepackage{bm}
\usepackage{amssymb}
\usepackage{mathrsfs}

\begin{document}
\title{Quantum limit in continuous quantum measurement}
\author{ChengGang Shao}
\email[]{cgshao@mail.hust.edu.cn}

\affiliation{Key Laboratory of Fundamental Physical Quantities Measurement of Ministry of Education, School of Physics, Huazhong University of Science and Technology, Wuhan 430074, People's Republic of China}

\date{\today}

\begin{abstract}
 An inequality about quantum noise is presented with the imprecise measurement theory, which is used to analyse the quantum limit in continuous quantum measurement. Different from the linear-response approach based on the quantum relation between noise and susceptibilities of the detector, we provide an explicit functional relation between quantum noise and reduction operator, and show a rigorous result: The minimum noise added by the detector in quantum measurement is precisely equal to the zero-point noise. This conclusion generalizes the standard Haus-Caves quantum limit for a linear amplifier. We also discuss the statistic characters of the back-action force in quantum measurement and show on how to reach the quantum limit.
\end{abstract}


\maketitle

\section{Introduction}
The subject of quantum measurement is as old as the very foundation of quantum mechanics. For a long time, the scheme proposed by von Neumann [1] is the main approach to the description of quantum measurements. According to this scheme, the quantum measurement is accompanied by an inevitable perturbation of the quantity that is canonically conjugate to the measured one. The magnitude of the minimum perturbation is constrained by the Heisenberg uncertainty relation, which has been shown in many ways [2-5]. However, this uncertainty relation for the continuous measurement has not been effectively discussed before, even lacks the explicit definitions of the noise and disturbance in reduction operator formalism. The study of the quantum perturbation or noise is worthwhile only in continuous quantum measurement with spectral representation. Although the measurements results can also be expressed in terms of a path integral formulation [6,7], it is usually complex and not suitable for the analytical calculation of the minimum quantum noise.

A simple method to calculate the minimum quantum noise in continuous measurement was developed by Braginsky [8] and Clerk [9,10] based on the linear-response theory. This approach treats the detector as a black box with an input port coupling to the signal source and an output port yielding the measurement results. The input port is characterized by an operator $\hat{F}$ with the noise spectral density $S_{FF}(\omega)$. And the output port is characterized by an operator $\hat{Z}$ with the noise spectral density $S_{ZZ}(\omega)$. By introducing the susceptibilities of the detector $\chi_{ZF}$ and $\chi_{FZ}$, this approach leads to the quantum noise inequality of the detector: a relation between the noise added to the output and the back-action noise feeding back to the signal source. For a dissipation-free system, the quantum noise inequality reduces to the simple form as [8,10]
\begin{eqnarray}
S_{FF}(\omega) S_{ZZ}(\omega)-|S_{ZF}(\omega)|^2 \geq && \frac{\hbar^2}{4}|\chi_{ZF}(\omega)-\chi_{FZ}(\omega)|^2 +\nonumber\\
&&\hbar \left| [\chi_{ZF}(\omega)-\chi_{FZ}(\omega)]\Im [S_{ZF}(\omega)]\right|.
\label{eq1}\
\end{eqnarray}
The minimum quantum noise is connected with an asymmetry between the forward and reverse transfer coefficients $\chi_{ZF}$ and $\chi_{FZ}$. For a symmetric system with $\chi_{ZF}=\chi_{FZ}$, the inequality (1) degenerates in to an identity and the noise may be zero. Obviously the inequality (1) can't be applied to the actual measurement process if one knows nothing about the constraints for the susceptibilities of the detector, such as the condition of simultaneous measurability [11], i.e. the susceptibilities that describe the response to a signal applied to the output port must vanish.

According to the von Neumann's postulate of reduction, the measurement process is irreversible. Once the measurement is finished and the information has been extracted from the measured object, the measured object cannot return to the initial state. Therefore, in quantum measurement, the noise added by a detector is unavoidable, and the right hand side in inequality (1) can't be zero. In this paper, we will develop a method to calculate the quantum noise $S_{FF}(\omega)$, $S_{ZZ}(\omega)$ and $S_{ZF}(\omega)$ of the detector in continuous quantum measurement. We devote attention to the relation between quantum noise and reduction operator, not the susceptibilities of the detector. Based on the imprecise measurement theory, we obtain the quantum noise inequality of the general detector as
\begin{eqnarray}
S_{FF}(\omega) S_{ZZ}(\omega)-|S_{ZF}(\omega)|^2 \geq  \frac{\hbar^2}{4},
\label{eq2}\
\end{eqnarray}
which is similar to the inequality (1) but an obvious difference is the absence of the susceptibilities. Instead, we will present an explicit functional relation between noise and reduction operator, which can be used to analyse the quantum limit in continuous measurement. Here we obtain a rigorous result: The minimum noise added by the detector is precisely equal to the noise arising from a zero temperature bath, which generalizes the standard Haus-Caves quantum limit for a linear amplifier.

Historically, Braginsky first discussed the quantum limit for the sensitivity of gravitational wave antennae. He used the inequality $S_{FF}S_{ZZ}\geq \hbar^2/4$ without the cross correlation term $S_{ZF}$, and obtained the 'standard quantum limit' (SQL) in the conventional measurement procedure [12,13]. Our calculation will show that the term $S_{ZF}$ plays an important role in beyond the SQL [14]. For the Hermitian reduction operator, $S_{ZF}$ is zero. While for the general case with non-Hermitian reduction operator, $S_{ZF}$ is nonzero. We will focus on the description of $S_{ZF}$ and show on how to achieve the quantum limit by choosing an appropriate non-Hermitian reduction operator.

In section 2 we define the quantum noise in measurement of an observable and the quantum noise of the detector. In section 3 we proved the quantum noise inequality of the detector, which is the heart of this paper. Section 4 is devoted to discuss the statistic characters of the back-action force with non-Hermitian reduction operator. In section 5 we compare description on the quantum noise between in our method and in linear-response method, and calculate the quantum limit in continuous measurement. The last section is a conclusion.

\section{Quantum noise in measurement of an observable}
We consider a quantum object described by the state $|\phi_{0}\rangle$, which interacts with the detector in measurement. Let $q$ be the observable and $\hat{q}$ its corresponding operator. According to the theory of quantum measurement, the measurement result of the observable $q$ is a stochastic process, which can be defined as
\begin{eqnarray}
\tilde{q}(t)\equiv q^{0}(t)+\delta q(t)+\int_{-\infty}^{t}\chi(t,t')F(t')dt'.
\label{eq3}\
\end{eqnarray}
The physical interpretation of this definition is simple. The first term $q^{0}(t)$ represents the intrinsic noise of the quantum object, which results from the probability interpretation of the wave function. The expectation value for the random variable $q^{0}(t)$ is $E[q^{0}(t)]=\langle\phi_{0}|\hat{q}(t)|\phi_{0}\rangle$. The second and last terms are the quantum noises in measurement process. $\delta q(t)$ and $F(t)$ represent the output and input noise of the detector, respectively. The measurement of an observable $q$ is accompanied by an inevitable perturbation of its conjugate quantity $F$. The magnitude of the minimum perturbation is given by the Heisenberg uncertainty relation. $\chi(t,t')$ is the generalized susceptibility of the measured object.

The purpose of this paper is to develop a method to calculate the total quantum noises added by the detector (the last two terms in Eq. (3)) using the description of imprecise measurement theory, and provide an exact expression of the total quantum noise as a function of the reduction operator. In most cases, the noise is usually expressed in frequency domain. In the following sections, we will derive the quantum noise inequality of the detector with spectral representation, and calculated the minimum quantum noise added by the detector.

\section{Quantum noise inequality of the detector}
\subsection{Measurement accuracy in imprecise measurement}
The general measurement theory can be quite simply expressed in terms of the density operator. Consider the instantaneous measurement of an observable $\hat{q}$. If the density operator of the measured object is initially $\hat{\rho }_{init}=|\phi_{0}\rangle \langle\phi_{0}|$, after the measurement the density operator has the form
\begin{eqnarray}
\hat {\rho }(\tilde {q}) =\frac{1}{w(\tilde {q})}\hat {\Omega }(\tilde
{q})\hat {\rho }_{init} \hat {\Omega }^\dag (\tilde {q}), \label{eq4}\
\end{eqnarray}
with the measurement outcome $\tilde {q}$, which has the following probability distribution
\begin{eqnarray}
w(\tilde{q})=\mathrm{Tr}(\hat{\Omega}^\dag(\tilde{q})\hat{\Omega}(\tilde{q})\hat{\rho}_{init}).
\label{eq5}\
\end{eqnarray}
$\hat{\Omega}^\dag(\tilde{q})\hat{\Omega}(\tilde{q})$ is called the 'effect' corresponding to the measurement or reduction operator $\hat{\Omega }(\tilde{q})$, known as Kraus operator. Eqs. (4) and (5) are the two fundamental postulates of the imprecise measurement [15]. In the following discussion, we shall focus on linear quantum-nondemolition measurement, in which the operator $\hat{\Omega }(\tilde{q})$ commutes with the measured quantity $\hat{q }$ [8]. In the $q$ representation, one can write the $\hat{\Omega }(\tilde{q})$ in the form
\begin{eqnarray}
\hat{\Omega}(\tilde{q})=\int_{-\infty}^{\infty}|q\rangle\Omega(\tilde{q}-q) \langle q|dq,
\label{eq6}\
\end{eqnarray}
where $\Omega(\tilde{q}-q)$ is some real or complex function that is normalized to unity   and satisfies the linear condition
\begin{eqnarray}
\int_{-\infty}^{\infty}(\delta\tilde{q})|\Omega(\delta\tilde{q})|^2|d(\delta\tilde{q})=0,
\label{eq7}\
\end{eqnarray}
with the difference  $\delta\tilde{q}=\tilde{q}-q$. The variance of $\delta\tilde{q}$ is given by
\begin{eqnarray}
(\Delta q)^2\equiv\int_{-\infty}^{\infty}(\delta\tilde{q})^2|\Omega(\delta\tilde{q})|^2| d(\delta\tilde{q}).
\label{eq8}\
\end{eqnarray}
The quantity $\Delta q$ is the rms measurement error.

\subsection{Sequences of discrete measurement}
Let us introduce a sequence of discrete measurement, which can easily transform to the continuous measurement by taking the appropriate continuous limit. Suppose that the observables $\hat{q}_1$, $\hat{q}_2, \cdots, \hat{q}_N$ are measured consecutively. The change of the state of the measured object in each measurement is described by Eq. (4). We rewrite it in the form as
\begin{eqnarray}
\hat{\rho}_{j-1} \rightarrow \hat{\rho}_{j} = \frac{1} {w(\tilde{q}_j|\tilde{q}_{j-1},\cdots,\tilde{q}_1)} \hat{\Omega }_j(\tilde{q}_j)\hat{\rho}_{j-1}\hat{\Omega }_j^\dag(\tilde{q}_j),j=1,2,\cdots N,
\label{eq9}\
\end{eqnarray}
where $w(\tilde{q}_j|\tilde{q}_{j-1},\cdots,\tilde{q}_1)$  is the conditional probability to obtain the measurement result $\tilde{q}_{j}$ in the $j$-th measurement if the previous $j$-1 measurement results are $\tilde{q}_{j-1},\cdots,\tilde{q}_{1}$. After the entire sequence of $N$ measurements is finished, the last state of the measured object is
\begin{eqnarray}
\hat{\rho}_{N}(\tilde{q}_1,\cdots\tilde{q}_N)=\frac{1}{w(\tilde{q}_{1},\cdots\tilde{q}_N)}\hat{\Omega }_N(\tilde{q}_N)\cdots \hat{\Omega }_1(\tilde{q}_1) \hat{\rho}_{init} \hat{\Omega}_1^\dag(\tilde{q}_1), \cdots \hat{\Omega}_N^\dag(\tilde{q}_N)
\label{eq10}\
\end{eqnarray}
with $w(\tilde{q}_{1},\cdots\tilde{q}_N)$ the joint probability distribution for the measured results.
\begin{eqnarray}
w(\tilde{q}_{1},\cdots,\tilde{q}_N)=\langle\hat{Z}_N^\dag(\tilde{q}_1, \cdots,\tilde{q}_N)\hat{Z}_N(\tilde{q}_1,\cdots,\tilde{q}_N)\rangle_{init},
\label{eq11}\
\end{eqnarray}
and $\hat{Z}_j(\tilde{q}_1, \cdots,\tilde{q}_j)\equiv\hat{\Omega}_j(\tilde{q}_j)\cdots \hat{\Omega}_1 (\tilde{q}_1)$. Here we introduce the notation $\langle\cdots\rangle_{init}=\mathrm{Tr}(\cdots \hat{\rho}_{init})$ [8]. Eq. (11) in principle gives a full description of the statistics of the results of a sequence of the measurements. However, the full probability distributions is not needed to know in practice, which is not applicable to the continuous limit. Instead, we usually only need to consider the first two moments of the distributions, i.e. the expectation values, the variance, and the cross correlations. Let us calculate them. Braginsky [8] has obtained some results of them for the case of the Hermitian operators $\hat{\Omega}_j$, but our calculation results apply to more general situations.

The expectation value for the $j$-th measurement is
\begin{eqnarray}
E[\tilde{q}_j] && \equiv \int_{\{\tilde{q}\}}w(\tilde{q}_1,\cdots\tilde{q}_N)\tilde{q}_j\cdot d\tilde{q}_1 d\tilde{q}_N \nonumber\\
&& =\int_{\{\tilde{q}\}} \langle \hat{Z}_{j-1}^\dag(\tilde{q}_1,\cdots\tilde{q}_{j-1})\hat{q}_{j} \hat{Z}_{j-1}(\tilde{q}_1,\cdots\tilde{q}_{j-1}) \rangle_{init} d\tilde{q}_1\cdots d\tilde{q}_{j-1}, \label{eq12}
\end{eqnarray}
where $\int_{\{\tilde{q}\}}$  means an integral over all possible results of the measurements. In order to see the physical meaning of the reduction operator $\hat{\Omega}_l$, we use the relation
\begin{eqnarray}
\hat{q}_j\hat{\Omega}_l=\hat{\Omega}_l\hat{q}_j+\frac{\partial \Omega_l(\tilde{q}_l-\hat{q}_l)}{\partial\hat{q}_l} \cdot i k_{jl},
\label{eq13}\
\end{eqnarray}
with $[\hat{q}_j,\hat{q}_l]=ik_{jl}$ to simplify Eq. (12). Linearity of the measurements implies that the commutators of all the measured observables are $c$-numbers. The $k_{jl}$  form an $N\times N$ matrix. By inserting Eq. (13) into (12) and repeating the manipulations, one arrives at
\begin{eqnarray}
E[\tilde{q}_j]=\langle\hat{q}_j\rangle_{init}+i\sum_{l=1}^{j-1}k_{jl}\int_{-\infty}^{\infty} \Omega_l^{\ast}\frac{d\Omega_l(q)}{dq}dq.
\label{eq14}\
\end{eqnarray}
The physical meaning of this relation is evident [8]: The mean value for the result of the $j$-th measurement is equal to the initial expectation value of the $j$-th observable, plus the sum of the perturbations of its mean in the first $j$-1 th measurements.

The second moment of the $j$-th measurement is given by
\begin{eqnarray}
E[\tilde{q}_j^2] && \equiv \int_{\{\tilde{q}\}}w(\tilde{q}_1,\cdots\tilde{q}_N)\tilde{q}_j^2 d\tilde{q}_1\cdots d\tilde{q}_N \nonumber\\
&& =\int_{\{\tilde{q}\}} \langle \hat{Z}_{j-1}^\dag [\hat{q}_{j}^2+(\Delta q_j)^2] \hat{Z}_{j-1} \rangle_{init} d\tilde{q}_1\cdots d\tilde{q}_{j-1},
\label{eq15}
\end{eqnarray}
with $\Delta q_j=[\int_{-\infty}^{\infty}q^2|\Omega_j(q)|^2 dq]^{\frac{1}{2}}$ the rms error of the $j$-th measurement. Using the relation (13) again, we can obtain the final form for the mean square result of the $j$-th measurement as
\begin{eqnarray}
E[\tilde{q}_j^2]=&&\langle\hat{q}_j^2\rangle_{init}+(\Delta q_j)^2+\sum_{l=1}^{j-1}k_{lj}^2 \int_{-\infty}^{\infty}|\Omega_{l}^{'}|^2dq+2i\sum_{l=1}^{j-1}k_{lj} \int\Omega_l^{\ast} \Omega_l^{'}dq \langle\hat{q}_j\rangle_{init} \nonumber\\
&&-\sum_{m=1}^{j-1}\sum_{n=1}^{j-1} k_{mj}k_{nj} \int\Omega_n^{\ast} \Omega_n^{'}dq \int\Omega_m^{\ast} \Omega_m^{'}dq+\sum_{m=1}^{j-1}(k_{mj}\int\Omega_m^{\ast} \Omega_m^{'}dq)^2.\nonumber\\
\label{eq16}
\end{eqnarray}

For the cross correlation of the results of the two different measurements, we skip the details of the calculation and generalize the formula (16) to the following form
\begin{eqnarray}
E[\tilde{q}_j\tilde{q}_l]=\langle\hat{q}_j \circ \hat{q}_l\rangle_{init}+(\Delta q_j)^2 \delta_{jl} +\sum_{m=1}^{l-1}k_{mj}k_{ml}[\int|\Omega_m^{'}|^2dq+(\int\Omega_m^{\ast}\Omega_m^{'}dq)^2] \nonumber\\ +\frac{i k_{lj}}{2} \int(\Omega_l^{\ast}\Omega_l^{'}-\Omega_l^{'\ast}\Omega_l)q dq + i\sum_{m=1}^{j-1}k_{mj} \int\Omega_m^{\ast} \Omega_m^{'}dq \langle\hat{q}_l\rangle_{init} \nonumber\\
+i\sum_{m=1}^{l-1}k_{ml} \int\Omega_m^{\ast} \Omega_m^{'}dq \langle\hat{q}_j\rangle_{init} -\sum_{m=1}^{j-1}\sum_{n=1}^{l-1} k_{mj}k_{nl} \int\Omega_n^{\ast} \Omega_n^{'}dq \int\Omega_m^{\ast} \Omega_m^{'}dq \label{eq17}
\end{eqnarray}
for $j\geq l$, with the symmetrized product $\hat{q}_j \circ \hat{q}_l$ defined by $\hat{q}_j \circ \hat{q}_l=(\hat{q}_j\hat{q}_l+\hat{q}_l\hat{q}_j)/2$.

From Eqs. (14) and (17), the covariant matrix can be directly calculated as
\begin{eqnarray}
B_{jl} & \equiv & E[\tilde{q}_j\tilde{q}_l]-E[\tilde{q}_j]E[\tilde{q}_l] \nonumber\\ & = & B_{jl}^{init}+(\Delta q_j)^2 \delta_{jl}+\sum_{m=1}^{l-1}k_{mj}k_{ml} [\int|\Omega_m^{'}|^2dq+(\int\Omega_m^{\ast}\Omega_m^{'}dq)^2] \nonumber\\ & & +\frac{i k_{lj}}{2} \int(\Omega_l^{\ast}\Omega_l^{'}-\Omega_l^{'\ast}\Omega_l)q dq,
\label{eq18}
\end{eqnarray}
for $j\geq l$, where the matrix $B_{jl}^{init}\equiv \langle\hat{q}_j \circ \hat{q}_l\rangle_{init}-\langle\hat{q}_j\rangle_{init} \langle\hat{q}_l\rangle_{init}$  is determined by the initial state of the measured object. The covariant matrix describes the statistical correlation of the measurement results. It is a discrete analog of the correlation function for random processes. Braginsky [8] has obtained the same results for the case of real functions $\Omega_j(q)$, or the Hermitian operators $\hat{\Omega}_j$. In the following, we will show that in order to achieve the quantum limit, $\Omega_j(q)$ should be taken the form of complex function.

\subsection{Quantum noise inequality in continuous measurements}
The continuous measurement can be regarded as the limiting case of a sequence of discrete measurements, when the time interval between the measurements tends to zero. Let us consider the discrete measurement of a variable $\hat{q}(t_j)$ in the Heisenberg picture. The time interval between two adjacent measurements is $\Delta t=t_j-t_{j-1}$. The rms error of the measurement at time $t_j$ is $\Delta q(t_j)=[\int_{-\infty}^{\infty}q^2|\Omega_j(q)|^2 dq]^{\frac{1}{2}}$. We also introduce the strength of the noise for the measurement $S_q(t_j)=\Delta t[\Delta q(t_j)]^2$. In the asymptotic limit $\Delta t\rightarrow 0$ and $\Delta q\rightarrow \infty$, the measurement-imprecision noise $S_q(t)$ stays finite as a continuous function of time
\begin{eqnarray}
S_q(t) \equiv \lim_{\Delta t \rightarrow 0} \Delta t \int_{-\infty}^{\infty}q^2|\Omega_t(q)|^2 dq,
\label{eq19}\
\end{eqnarray}
replacing $\Omega_j$ by $\Omega_t$ for continuous measurement. Similarly, the strength of the continuous perturbance is defined by
\begin{eqnarray}
S_F(t) \equiv \lim_{\Delta t \rightarrow 0} \frac{\hbar^2}{\Delta t} [\int|\Omega_t^{'}|^2dq+(\int\Omega_t^{\ast}\Omega_t^{'}dq)^2].
\label{eq120}\
\end{eqnarray}

The covariant function for the continuous measurement can be obtained as follows, by replacing the summation in Eq. (18) with an integral,
\begin{eqnarray}
B(t,t') =&& B^{init}(t,t')+S_q(t)\delta(t-t')+\int_{-\infty}^{t'}\chi(t,t'')\chi(t',t'') S_F(t'')dt''  \nonumber\\ && - \frac{i \hbar}{2} \chi(t,t') \int(\Omega_{t'}^{\ast}\Omega_{t'}^{'}-\Omega_{t'}^{'\ast}\Omega_{t'})q dq,
\label{eq21}
\end{eqnarray}
for $t\geq t'$ , where, the generalized susceptibility $\chi(t,t')$ of the measured object is defined by the self-commutator of any generalized coordinate as
\begin{eqnarray}
\chi(t,t')=\frac{i}{\hbar}[\hat{q}(t),\hat{q}(t')]\Theta(t-t')=-\frac{1}{\hbar}k(t,t')\Theta(t-t')
\label{eq22}\
\end{eqnarray}
with the step function $\Theta(t)$.
\begin{eqnarray}
B^{init}(t,t')\equiv \langle \hat{q}(t)\circ \hat{q}(t') \rangle_{init}-\langle \hat{q}(t)\rangle_{init} \langle \hat{q}(t') \rangle_{init},
\label{eq23}\
\end{eqnarray}
is the 'unperturbed correlation function' that describes the statistical properties of the measured quantity in the absence of any influence from the detector. $S_F(t)$ in Eq. (21) characters the strength of the random force that the measuring device or detector exerts on the measured object. The correlation function of this back-action force is
\begin{eqnarray}
B_F(t,t')\equiv S_F(t)\delta(t-t'),
\label{eq24}\
\end{eqnarray}
in continuous measurement, as well as the correlation function of the noise of the measuring device
\begin{eqnarray}
B_q(t,t')\equiv S_q(t)\delta(t-t').
\label{eq25}\
\end{eqnarray}
The last term in Eq. (21) describes the cross correlation between the measurement noise and the back-action force,
\begin{eqnarray}
B_{qF}(t,t')\equiv S_{qF}(t)\delta(t-t')=-\frac{i \hbar}{2} \int(\Omega_{t}^{\ast}\Omega_{t}^{'}-\Omega_{t}^{'\ast}\Omega_{t})q dq \delta (t-t').
\label{eq26}\
\end{eqnarray}

Eqs. (19), (20) and (26) imply the following uncertain relation of general quantum measurement
\begin{eqnarray}
S_q(t)S_F(t)-S_{qF}^2(t)\geq \frac{\hbar^2}{4},
\label{eq27}\
\end{eqnarray}
which is equivalent to the general form of the Heisenberg uncertainty relation for two observables $\hat{A}$ and $\hat{B}$ [16]
\begin{eqnarray}
\langle \Delta \hat{A}^2 \rangle \langle \Delta \hat{B}^2 \rangle \geq \frac{1}{4} \langle \{\Delta \hat{A},\Delta \hat{B}\} \rangle^2 +\frac{1}{4} \langle [\Delta \hat{A},\Delta \hat{B}] \rangle^2,
\label{eq28}\
\end{eqnarray}
with $\Delta \hat{A}\equiv \hat{A}-\langle\hat{A}\rangle$ and $\Delta \hat{B}\equiv \hat{B} - \langle \hat{B} \rangle$. Thus Eq. (27) is directly obtained by setting $\hat{A}=q$ and $\hat{B}=-i\hbar \partial_q$ into Eq. (28) for the state vector or wave function $\Omega_t(q)$ in $q$ representation. Eq. (27) is the general quantum noise inequality in continuous measurement, which remains valid when the accuracy of the measurement varies with the time. The equality is achieved in Eq. (27) only for the complex Gauss function
\begin{eqnarray}
\Omega_t(q)=\lim_{\Delta t \rightarrow 0} \frac{1}{[2\pi(\Delta q)^2]^{1/4}} \exp\left(-\frac{1-i2S_{qF}(t)/\hbar}{4(\Delta q)^2} q^2 - \frac{i\Delta t \bar{F}(t)q}{\hbar}\right),
\label{eq29}\
\end{eqnarray}
with any real function
\begin{eqnarray}
\bar{F}(t)=\lim_{\Delta t \rightarrow 0} \frac{i\hbar}{\Delta t} \int_{-\infty}^{\infty} \Omega_t^{\ast} \frac{d\Omega_t(q)}{dq}dq.
\label{eq30}\
\end{eqnarray}
$\bar{F}(t)$ represents the mean value of the random back-action force that the detector exerts on the measured object.

\subsection{Spectral representation of the quantum noise in continuous measurement}
In above discussion, we have assumed that the noise of imprecise measurement is uncorrelated for different time $t$ and $t'$ as expressed in Eq. (25). We now consider a different situation, where $B_q(t,t')$ depends only on the time difference $t-t'$. Thus, the spectral method can be used in this stationary system to simplify calculations. For such a stationary system, both the accuracy of the monitoring and the strength of the back perturbance are constant. The susceptibility and the correlation functions for measured quantity also depend only on the time difference $t-t'$. This permits us to represent the measured quantity by the Fourier transform
\begin{eqnarray}
S_{\tilde{q}\tilde{q}}(\omega) \equiv \int_{-\infty}^{\infty} B(\tau)e^{i\omega\tau}d\tau, S_{qq}^{init}(\omega) \equiv \int_{-\infty}^{\infty} B^{init}(\tau)e^{i\omega\tau}d\tau,
\label{eq31}\
\end{eqnarray}
with time difference $\tau=t-t'$, as well as the same definition for the susceptibility $\chi(\omega)$. The Fourier transform of $B_q(t,t')$ is
\begin{eqnarray}
S_{qq}(\omega)\equiv\int_{-\infty}^{\infty} B_q(\tau)e^{i\omega\tau}d\tau,
\label{eq32}\
\end{eqnarray}
as well as for $S_{FF}(\omega)$ and $S_{qF}(\omega)$. Then Eqs. (21) and (27) can be transformed to the frequency domain
\begin{eqnarray}
S_{\tilde{q}\tilde{q}}(\omega) =S_{qq}^{init}(\omega) + S_{qq}(\omega)+|\chi(\omega)|^2S_{FF}(\omega) + 2 \Re[\chi(\omega)]S_{qF}(\omega), \label{eq33}\\
S_{qq}(\omega) S_{FF}(\omega)-S_{qF}^2(\omega) \geq \frac{\hbar^2}{4},
\label{eq34}
\end{eqnarray}
respectively. These results based on the description of imprecise measurement are different to Eq. (1) given by the linear-response detector description [10,8], where the susceptibilities of the detector are introduced in stead of the reduction operator $\hat{\Omega}$.

\section{Stochastic master equation and back-action force with non-Hermitan reduction operator}
We have pointed out the minimum noise of the detector can be achieved for the non-Hermitian reduction operator, or the complex reduction function as in Eq. (29). In this section we further consider the statistic characteristics of the back-action force for the general non-Hermitian reduction operator. There are a simple model of an indirect measurement developed by Caves [6], which is very suitable to calculate the relation between the stochastic back-action force and the reduction function.

The measurement in this model is a two-step process that leads naturally to the standard description of the imprecise measurement. In the first step the measured object interacts with a detector that has been prepared in some special initial quantum state, producing a correlation such that the state of the detector acquires some information about the measured object. The second step applies the reduction postulate to obtain the readout of the detector.

In the model, the complete Hamiltonian of the joint system (measured object plus the detector) is given by [6,17]
\begin{eqnarray}
\hat{H}=\hat{H}_{0}-\sum_{j=1}^{N}\delta(t-j \Delta t) \hat{q} \hat{P}_j,
\label{eq35}\
\end{eqnarray}
where $\hat{H}_{0}$ is the free Hamiltonian of the system. The interaction between the measured object and the detector is described by $\delta(t-j \Delta t) \hat{q} \hat{P}_j$  with the object variable  $\hat{q}$ and the detector variable $\hat{P}_j\equiv \hat{F}_j\Delta t$  for the $j$-th measurement. Operator $\hat{F}_j$  indicates the back-action force in the measurement process. In each measurement the detector is prepared in the state $|\psi_j\rangle$. The measurement result $\tilde{q}_j$ is determined by change of the state of the detector according to the reduction postulate, which is simply marked by $|\tilde{q}_j\rangle$. We rewrite Eq. (9), the object's state after each measurement, in the form
\begin{eqnarray}
\hat{\rho}_{j-1} \rightarrow \hat{\rho}_{j} = \frac{1} {w(\tilde{q}_j)} \langle \tilde{q}_j|\exp[\frac{i}{\hbar}\hat{q}_j \hat{F}_j \Delta t]|\psi_{j-1} \rangle \hat{\rho}_{j-1} \langle \psi_{j-1}|\exp[\frac{-i}{\hbar}\hat{q}_j \hat{F}_j \Delta t]|\tilde{q}_j \rangle, \nonumber\\
\label{eq36}\
\end{eqnarray}
in the Heisenberg picture with $\hat{q}_j\equiv \hat{q}(t_j)=e^{i\hat{H}_0 \Delta t/\hbar}\hat{q}_{j-1} e^{-i\hat{H}_0 \Delta t/\hbar}$, as well as $\hat{F}_j\equiv \hat{F}(t_j)$. The corresponding reduction operator
\begin{eqnarray}
\hat{\Omega}(\tilde{q}_j)= \langle \tilde{q}_j|\exp[\frac{i}{\hbar}\hat{q}_j \hat{F}_j \Delta t]|\psi_{j-1} \rangle = \Omega(\tilde{q}_j-\hat{q}_j).
\label{eq37}\
\end{eqnarray}
The normalization factor $w(\tilde{q}_j)$ can be transformed into the exponential form by defining $\bar{q}_j$, which satisfies $|\Omega(\tilde{q}_j-\bar{q}_j)|^2=w(\tilde{q}_j)$, then Eq. (36) becomes
\begin{eqnarray}
\hat{\rho}_{j-1} \rightarrow \hat{\rho}_{j} = && \frac{1} {w(\tilde{q}_j)} \langle \tilde{q}_j| \exp[i \bar{q}_j \hat{F}_j\Delta t/\hbar] \exp[i (\hat{q}_j-\bar{q}_j) \hat{F}_j\Delta t/\hbar]|\psi_{j-1} \rangle \hat{\rho}_{j-1} \nonumber\\
&&\times \langle \psi_{j-1}|\exp[-i(\hat{q}_j-\bar{q}_j) \hat{F}_j\Delta t/\hbar] \exp[-i \bar{q}_j \hat{F}_j\Delta t/\hbar].
\label{eq38}\
\end{eqnarray}

In order to see the effect of the random back-action force, we use the relation $e^BAe^{-B}=A+[B,A]+\cdots$ to expand the Eq. (38) in a series as
\begin{eqnarray}
\hat{\rho}_{j}=&& \hat{\rho}_{j-1} + \frac{i \Delta t} {\hbar} [(\hat{q}_j-\bar{q}_j)F_j,\hat{\rho}_{j-1}] - \frac{i (\Delta t)^2} {2\hbar^2} [(\hat{q}_j+\bar{q}_j)F_j,[(\hat{q}_j-\bar{q}_j)F_j,\hat{\rho}_{j-1}]] \nonumber\\
&&+o[(\Delta t)^3],
\label{eq39}\
\end{eqnarray}
where, we have introduced the classical back-action force $F_j$  at $j$-th measurement. The explicit form of $F_j$ satisfies the following equations
\begin{eqnarray}
F_{j}\hat{\rho}_{j-1}&=\frac{i\hbar}{w(\tilde{q}_j)\Delta t} \Omega_{j}^{\ast}(\tilde{q}_j) \Omega_{j}^{'}(\tilde{q}_j) \hat{\rho}_{j-1}, \nonumber\\
\hat{\rho}_{j-1}F_{j}&=\frac{-i\hbar}{w(\tilde{q}_j)\Delta t} \Omega_{j}^{\ast '}(\tilde{q}_j) \Omega_{j}(\tilde{q}_j) \hat{\rho}_{j-1},
\end{eqnarray}
which mean that $F_j$  and $\hat{\rho}_{j-1}$  are noncommutative. But the expectation of  $F_j$  and $\hat{\rho}_{j-1}$ are commutative, $E[F_j\hat{\rho}_{j-1}]=E[\hat{\rho}_{j-1}F_j]$. The mean value of the back-action force is
\begin{eqnarray}
E[F_j]=\frac{i\hbar}{\Delta t} \int_{-\infty}^{\infty} \Omega_{j}^{\ast}(q)\Omega_{j}^{'}(q)dq = \frac{-i\hbar}{\Delta t} \int_{-\infty}^{\infty} \Omega_{j}^{\ast '}(q)\Omega_{j}(q)dq,
\label{eq41}\
\end{eqnarray}
which is equal to Eq. (30) in continuous measurement limit. If the reduction function $\Omega_{j}(q)$ are real, in other words, the operator $\hat{\Omega}_j$ is Hermitan, then we have $F_{j}\hat{\rho}_{j-1} = -\hat{\rho}_{j-1}F_{j}$, which implies the means value of the back-action force is zero. Similarly, the square terms of $F_j$ in Eq. (39) are
\begin{eqnarray}
F_{j}\hat{\rho}_{j-1}F_{j} && = \frac{\hbar^2}{w(\tilde{q}_j)(\Delta t)^2} |\Omega_{j}^{'}|^2 \hat{\rho}_{j-1}, \nonumber\\
F_{j}F_{j}\hat{\rho}_{j-1} && = \frac{-\hbar^2}{w(\tilde{q}_j)(\Delta t)^2} \Omega_{j}^{\ast}(\tilde{q}_j) \Omega_{j}^{''}(\tilde{q}_j) \hat{\rho}_{j-1}, \nonumber\\
\hat{\rho}_{j-1}F_{j}F_{j}&&=\frac{-\hbar^2}{w(\tilde{q}_j)(\Delta t)^2} \Omega_{j}^{\ast ''}(\tilde{q}_j) \Omega_{j}(\tilde{q}_j) \hat{\rho}_{j-1},
\end{eqnarray}
which have the same expectation with
\begin{eqnarray}
E[F_j^2]=\frac{\hbar^2}{(\Delta t)^2} \int_{-\infty}^{\infty} |\Omega_{j}^{'}|^2 dq.
\label{eq43}\
\end{eqnarray}
The corresponding variance is
\begin{eqnarray}
D[F_j^2] &&\equiv E[F_j^2]-E[F_j]^2 \nonumber\\
&&=\frac{\hbar^2}{(\Delta t)^2}\left [\int |\Omega_{j}^{'}|^2 dq + (\int \Omega_{j}^{\ast}\Omega_{j}^{'}dq)^2 \right ]=\frac{S_F(t_j)}{\Delta t}.
\label{eq44}\
\end{eqnarray}

The stochastic master equation can be written in the first order $\Delta t$ as
\begin{eqnarray}
\frac{\hat{\rho}_{j}-\hat{\rho}_{j-1}}{\Delta t} = -\frac{\int_{-\infty}^{\infty} |\Omega_{j}^{'}|^2 dq} {2(\Delta t)} [\hat{q}_j,[\hat{q}_j,\hat{\rho}_{j-1}]] + \frac{i}{\hbar}[(\hat{q}_j-\bar{q}_j)F_{j},\hat{\rho}_{j-1}].
\label{eq45}\
\end{eqnarray}
The classical back-action force $F_j$ is defined by Eqs. (40) for the general complex reduction function. Only in the case of real function $\Omega_j(q)$, we have $F_{j}\hat{\rho}_{j-1}=-\hat{\rho}_{j-1}F_{j}$, and above equation can be further written as the general Ito equation [15,18-21]
\begin{eqnarray}
d\hat{\rho}= -\frac{S_F(t)} {2\hbar^2}[\hat{q}(t),[\hat{q}(t),\hat{\rho}]]dt + \frac{S_F(t)^{1/2}} {\hbar}\{\hat{q}(t)-\bar{q}(t),\hat{\rho}\}dW,
\label{eq46}\
\end{eqnarray}
with $dW$ a differential of a real Wiener noise and $\bar{q}(t)=\mathrm{Tr}[\hat{q}(t)\hat{\rho}]$.

\section{Quantum limit in continuous measurement}
Although the minimum quantum noise added by the detector can be directly obtained according to Eqs. (33) and (34), it is meaningful to compare the differences between Eq. (34) in our approach and Eq. (1) in linear-response theory. This section will give a simple review of the quantum noise in linear-response detector description, and show how to constrain the susceptibilities of the detector, as well as the reduction function, to achieve the quantum limit.
\subsection{Description with the linear-response theory}
The linear-response theory does not use any notion of state collapse, and emphasizes the idea that a detector has the intrinsic noise limited by the Heisenberg uncertainty relation, which is constrained by the susceptibilities of the detector. Consider the joint system for the measured object and the detector in linear-response theory. The detector has both an input port, characterized by an operator $\hat{F}$, and an output port, characterized by an operator $\hat{Z}$. The output $\hat{Z}$ is the quantity at the read-out of the output of the detector. The input operator $\hat{F}$ is the detector quantity that directly couples to the input signal and that causes a back-action disturbance of the measured object. By applying the perturbation theory with linear approximation, one can derive the Heisenberg operators of the output variable $\hat{Z}$ of the detector as
\begin{eqnarray}
\hat{Z}_{pert}(t)= \hat{Z}(t)+\int_{-\infty}^{t}\chi_{ZF}(t,t')\hat{q}_{pert}(t')dt',
\label{eq47}\
\end{eqnarray}
with the variable $\hat{q}_{pert}$ of the measured object
\begin{eqnarray}
\hat{q}_{pert}(t)= \hat{q}(t)+\int_{-\infty}^{t}\chi(t,t')\hat{F}(t')dt'.
\label{eq48}\
\end{eqnarray}
$\hat{Z}(t)$ and $\hat{q}(t)$ are the operators in the absence of the interaction, which evolve under the free Hamiltonian. Here, we have assumed that the susceptibilities $\chi_{FZ}$, $\chi_{ZZ}$ and $\chi_{FF}$ are zero, only the susceptibility
\begin{eqnarray}
\chi_{ZF}(t,t')=\frac{i}{\hbar}[\hat{Z}(t),\hat{F}(t')]\Theta(t-t')
\label{eq49}\
\end{eqnarray}
nonzero. Here the asymmetry between the input and output of the detector is necessary because the measurement process is irreversible, which is a disadvantage of the linear-response theory.

By making the transition from the detector variable $\hat{Z}_{pert}(t)$ to the variable $\tilde{q}(t)$ of the measured object, we obtain the result in continuous measurement [8]
\begin{eqnarray}
\tilde{q}(t) && \equiv \int_{-\infty}^{t}\chi_{ZF}^{-1}(t,t')\hat{Z}_{pert}(t')dt' \nonumber\\
&& =\int_{-\infty}^{t}\chi_{ZF}^{-1}(t,t')\hat{Z}(t')dt' +\hat{q}(t)+ \int_{-\infty}^{t}\chi(t,t')\hat{F}(t')dt',
\label{eq50}\
\end{eqnarray}
where $\chi_{ZF}^{-1}$ is the inverse of $\chi_{ZF}$
\begin{eqnarray}
\int_{t'}^{t}\chi_{ZF}^{-1}(t,t'')\chi_{ZF}(t'',t')dt''= \delta (t-t').
\label{eq51}\
\end{eqnarray}
The simultaneous measurability requires the output signal $\tilde{q}(t)$, in effect, a $c$-number and not an operator [8,11].

In order to describe the strength of the random noise caused by the detector, we consider the noise in Fourier domain as
\begin{eqnarray}
\tilde{q}(\omega)=\lim_{T\rightarrow\infty}\frac{1}{T^{1/2}} \int_{0}^{T}\left(\tilde{q}(t)-E[\tilde{q}(t)] \right)e^{i\omega t}dt.
\label{eq52}\
\end{eqnarray}
Here the expectation $E[\tilde{q}(t)]$ has been subtracted from $\tilde{q}(t)$, and the same is true in the definition for $\hat{Z}(\omega)$ and $\hat{F}(\omega)$. $T$ is the sampling time. For stationary systems, where the susceptibility depends only on the time difference $t-t'$, the spectrum density of the total measured noise for the observable $q$ is
\begin{eqnarray}
S_{\tilde{q}\tilde{q}}(\omega)&& \equiv E[\tilde{q}(\omega)\tilde{q}(\omega)^{\dag}] \nonumber\\ &&=S_{qq}^{init}(\omega)+\frac{S_{ZZ}(\omega)}{|\chi_{ZF}(\omega)|^2}+|\chi_{qq}(\omega)|^2 S_{FF}(\omega) +2\Re \left [\chi(\omega)^{\ast} \frac{S_{ZF}(\omega)}{\chi_{ZF}(\omega)}\right ], \nonumber\\
\label{eq53}\
\end{eqnarray}
for sufficiently large $T$.

Without any information of the detector's Hamiltonian or state, one can get a general quantum constraint on its noise properties [8]
\begin{eqnarray}
S_{FF}(\omega) S_{ZZ}(\omega)-|S_{ZF}(\omega)|^2 \geq \frac{\hbar^2}{4}|\chi_{ZF}(\omega)|^2 + \hbar \left| \Im [\chi_{ZF}(\omega)^{\ast}S_{ZF}(\omega)] \right|.
\label{eq54}\
\end{eqnarray}
Comparing with discussion in the section 3, if we take the gain or the susceptibility $\chi_{ZF}(\omega)=1$ and $S_{ZF}(\omega)\in R $, Eqs. (53) and (54) will return to Eqs. (33) and (34), respectively, by changing the subscript $Z$ to $q$.

\subsection{Quantum limit of the detector}
 At last we turn to calculate the minimum quantum noise added by the general detector. It is useful to take the total measured noise in the output of the detector, Eq. (33) or the spectral representation of Eq. (3), in the form as [10]
\begin{eqnarray}
S_{\tilde{q}\tilde{q}}(\omega) \equiv S_{qq}^{init}(\omega)+S_{qq,add}(\omega).
\label{eq55}\
\end{eqnarray}
$S_{qq}^{init}$ represents the intrinsic noise of the measured object. For the equilibrium system, the fluctuation-dissipation theorem gives $S_{qq}^{init}=\hbar\coth(\hbar\omega/2k_BT_B) |\Im [\chi(\omega)]|$ at temperature $T_B$. $S_{qq,add}$ represents the added noise due to the detector. It has contributions from the detector's intrinsic output noise $S_{qq}(\omega)$ corresponding to the accuracy of measurement, from the detector's back-action noise $S_{FF}(\omega)$, and from the correlative term $S_{qF}(\omega)$.
\begin{eqnarray}
S_{qq,add}(\omega) =S_{qq}(\omega) + |\chi(\omega)|^2 S_{FF}(\omega)+2\Re [\chi(\omega)]S_{qF}(\omega).
\label{eq56}\
\end{eqnarray}
Defining $\phi(\omega)=\arg\chi(\omega)$, we thus have the bound
\begin{eqnarray}
S_{qq,add}(\omega) \geq 2|\chi(\omega)| \left[S_{qq}(\omega)^{1/2}S_{FF}(\omega)^{1/2}+\cos[\phi(\omega)]S_{qF}(\omega)\right],
\label{eq57}\
\end{eqnarray}
where the minimum value at frequency $\omega$ is achieved when
\begin{eqnarray}
S_{qq}(\omega) = |\chi(\omega)|^2 S_{FF}(\omega).
\label{eq58}\
\end{eqnarray}
Using the quantum noise constraint of Eq. (34) to further bound $S_{qq,add}(\omega)$, we obtain
\begin{eqnarray}
\min S_{qq,add}(\omega) = \hbar |\Im [\chi(\omega)]|,
\label{eq59}\
\end{eqnarray}
which tells that at each frequency the minimum noise added by the detector is precisely equal to the noise arising from a zero temperature bath.

The minimum $S_{qq,add}(\omega)$ is achieved only for the reduction operator with the Gauss function $\Omega$ defined in Eq. (29), or equivalently
\begin{eqnarray}
\Omega(q)=\frac{1}{[2\pi(\Delta q)^2]^{1/4}} \exp\left (-\frac{1+i \cot\phi(\omega)}{4(\Delta q)^2} q^2 - \frac{i\Delta t \bar{F}(\omega)q}{\hbar}\right).
\label{eq60}\
\end{eqnarray}
Noting that $\lim_{\Delta t\rightarrow 0}\Delta t(\Delta q)^2=S_{qq}(\omega)$, the corresponding measurement-imprecision noise, back-action noise and the cross correlation noise of the detector are
\begin{eqnarray}
S_{qq}(\omega) =\left|\frac{\hbar \chi(\omega)}{2 \sin\phi(\omega)}\right|,
\label{eq61}\
\end{eqnarray}
\begin{eqnarray}
S_{FF}(\omega)=\frac{\hbar^2}{4S_{qq}(\omega)}[1+\cot^2\phi(\omega)],
\label{eq62}\
\end{eqnarray}
and
\begin{eqnarray}
S_{qF}(\omega) =- \frac{\hbar}{2} \cot\phi(\omega),
\label{eq63}\
\end{eqnarray}
respectively.

\section{Conclusion}
In this paper we have obtained the quantum noise inequality of the general detector, which is used to analyse the quantum limit in continuous measurement. The measurement-imprecision noise, the strength of the random back force and the cross correlation noise satisfy the Heisenberg uncertainty relation (27), or equivalently for Eq. (34) in the stationary case. All the noises is determined by the measurement or reduction operator $\hat{\Omega}$, which is the amplitude for the detector to have evolved in each instantaneous measurement. The minimum quantum noise due to the detector can be achieved by choosing the non-Hermitian operator $\hat{\Omega}$ with the complex Gauss function $\Omega$ as in Eq. (60).

The minimum noise added by the detector is precisely equal to the noise arising from a zero temperature bath. This conclusion generalizes the result that given by Haus-Caves. Haus and Caves studied the quantum noise of the linear amplifiers [22,23]. They showed that, in the limit of a large power gain, the minimum noise added by the detector is precisely equal to the noise arising from a zero temperature bath. Equally, if a detector does not amplify (i.e., the power gain is unity), it need not produce any added noise, which was also clarified by Clerk [9,10]. However, without introduce the susceptibilities of the detector, we also obtain the quantum noise inequality in continuous limit, and show that the minimum noise of a detector in quantum measurement can't be zero. According to the von Neumann's postulate of reduction, the measurement process is irreversible. Once the measurement is finished and the information has been extracted, the measured object cannot return to the pre-measurement state. Therefore, the noise added by a detector is unavoidable, which must be at least as large as the zero-point noise.

\begin{acknowledgments}
This work was supported by the National Basic Research Program of China under Grant No.2010CB832802.
\end{acknowledgments}

\end{document}